 \documentclass[final, 5p, times, twocolumn]{elsarticle}

\usepackage{amssymb}

\journal{Astroparticle Physics}

\begin{document}

\begin{frontmatter}



\title{Searches for very high energy gamma rays from blazars with CANGAROO-III telescope in 2005--2009}


\author[1]{Y.~Mizumura\corref{a}}  \ead{mizumura@tkikam.sp.u-tokai.ac.jp}
\author[1]{J.~Kushida\corref{a}}   \ead{kushida@tkikam.sp.u-tokai.ac.jp}
\author[1]{K.~Nishijima\corref{a}} \ead{kyoshi@tkikam.sp.u-tokai.ac.jp}
\author[2]{G.~V.~Bicknell}
\author[3]{R.~W.~Clay}
\author[4]{P.~G.~Edwards}
\author[5]{S.~Gunji}
\author[6]{S.~Hara}
\author[7]{S.~Hayashi}
\author[8]{S.~Kabuki}
\author[7]{F.~Kajino}
\author[1]{A.~Kawachi}
\author[9]{T.~Kifune}
\author[10]{R.~Kiuchi}
\author[1]{K.~Kodani}
\author[11]{Y.~Matsubara}
\author[12]{T.~Mizukami}
\author[13]{Y.~Mizumoto}
\author[14]{M.~Mori}
\author[15]{H.~Muraishi}
\author[6]{T.~Naito}
\author[9]{M.~Ohishi}
\author[3]{V.~Stamatescu}
\author[3]{D.~L.~Swaby}
\author[12]{T.~Tanimori}
\author[3]{G.~Thornton}
\author[5]{F.~Tokanai}
\author[9]{T.~Toyama}
\author[16]{S.~Yanagita}
\author[16]{T.~Yoshida}
\author[9]{T.~Yoshikoshi}

\address[1]{ Department of Physics, Tokai University, Hiratsuka, Kanagawa 259-1292, Japan}
\address[2]{ Research School of Astronomy and Astrophysics, Australian National University, ACT 2611, Australia} 
\address[3]{ School of Chemistry and Physics, University of Adelaide, SA 5005, Australia}
\address[4]{ Australia Telescope National Facility, CSIRO, Epping, NSW 2121, Australia}
\address[5]{ Department of Physics, Yamagata University, Yamagata, Yamagata 990-8560, Japan}
\address[6]{ Faculty of Management Information, Yamanashi Gakuin University, Kofu, Yamanashi 400-8575, Japan}
\address[7]{ Department of Physics, Konan University, Kobe, Hyogo 658-8501, Japan}
\address[8]{ Department of Radiation Oncology, Tokai University, Isehara, Kanagawa 259-1193, Japan}
\address[9]{ Institute for Cosmic Ray Research, University of Tokyo, Kashiwa, Chiba 277-8582, Japan}
\address[10]{ Institute of Particle and Nuclear Studies, High Energy Accelerator Research Organization, Tsukuba, Ibaraki 305-0801, Japan}
\address[11]{ Solar-Terrestrial Environment Laboratory,  Nagoya University, Nagoya, Aichi 464-8602, Japan}
\address[12]{ Department of Physics, Kyoto University, Sakyo-ku, Kyoto 606-8502, Japan}
\address[13]{ National Astronomical Observatory of Japan, Mitaka, Tokyo 181-8588, Japan}
\address[14]{ Department of Physics, College of Science and Engineering, Ritsumeikan University, Kusatsu, Shiga 525-8577, Japan}
\address[15]{ School of Allied Health Sciences, Kitasato University, Sagamihara, Kanagawa 228-8555, Japan}
\address[16]{ Faculty of Science, Ibaraki University, Mito, Ibaraki 310-8512, Japan}

\cortext[a]{ Corresponding authors. }

\begin{abstract}
We have searched for very high energy (VHE) gamma rays from four blazars using the CANGAROO-III imaging atmospheric Cherenkov telescope.
We report the results of the observations of
H$~$2356$-$309, PKS$~$2155$-$304, PKS$~$0537$-$441, and 3C$~$279, performed from 2005 to 2009,
applying a new analysis to suppress the effects of the position dependence of Cherenkov images in the field of view.
No significant VHE gamma ray emission was detected from any of the four blazars.
The GeV gamma-ray spectra of these objects were obtained by analyzing Fermi/LAT archival data.
Wide range (radio to VHE gamma-ray bands)
spectral energy distributions (SEDs) including CANGAROO-III upper limits,
GeV gamma-ray spectra, and archival data, even though they are non-simultaneous, are discussed
using a one-zone synchrotron self-Compton (SSC) model in combination with a external Compton (EC) radiation.
The HBLs (H$~$2356$-$309 and PKS$~$2155$-$304) can be explained by a simple SSC model,
and PKS$~$0537$-$441 and 3C$~$279 are well modeled by a combination of SSC and EC model.
We find a consistency with the blazar sequence in terms of strength of magnetic field and component size.
\end{abstract}


\end{frontmatter}

\renewcommand\thefootnote{\alph{footnote}}


\section{Introduction}
\label{}
Blazars are a sub-class of active galactic nuclei (AGN), with 
spectral energy distributions (SEDs) that are characterized by double-peaked nonthermal emission
which extends from radio to gamma rays.
In addition, they are also characterized by rapid flux variability in all bands, polarization in radio and optical bands, and highly collimated relativistic jets.
AGNs have been classified into various classes according to the observed features,
which can be understood by differences of the angle between the jet and our line of sight dominantly, 
and also circumnuclear obscuration and relativistic beaming are candidates of deciding factor \citep{urry_95}.
Blazars include BL Lacertae objects (BL Lacs) and flat spectrum radio quasars (FSRQs),
which have jets within a few degrees of the line of sight.
Following the Whipple group's discovery of the first TeV gamma-ray blazar Mkn 421 \citep{punch_92},
blazars have became one of the most interesting class of object for VHE gamma-ray astronomy.

Recently, the AGILE \citep{tavani_08} and Fermi \citep{atwood_09} gamma-ray space satellites have been launched,
the Fermi/LAT detector has sensitivities about factor 30 enhanced from EGRET \cite{thompson_93}.
Combining simultaneous observational data from these satellites with VHE gamma-ray data from imaging atmospheric Cherenkov telescopes (IACTs)
allow us to study multiwavelength behavior in a wide range.

H$~$2356$-$309 is a high-frequency peaked BL Lac (HBL) at a redshift of $z=0.165$ \citep{falomo_91},
which was discovered in X-rays by the UHURU satellite \citep{forman_78}.
Some authors classified this object as "extreme HBL" \citep{costamante_01, ghisellini_02} based on its synchrotron peak at extremely high frequency.
In 2004, the H.E.S.S.\ group discovered this object in the VHE gamma-ray band above 0.2\,TeV
at the $10\sigma$ level of significance \citep{aharonian_06a, aharonian_06b},
as predicted by \citet{costamante_02}.
The observed VHE gamma-ray spectrum between 0.2\,TeV and 1.3\,TeV is consistent with a power-law with photon index of $\Gamma=3.1$.
H.E.S.S.\ observations were additionally made from 2005 to 2007 and they found little flux variability on the time scale of a few years \citep{abramowski_10}.
The CANGAROO-III telescope was pointed to H$~$2356$-$309 in 2005, but no evidence of VHE gamma-ray emissions above 750\,GeV was found \citep{nishijima_09}.
The flux upper limits are about one order of magnitude higher than the H.E.S.S. reported spectra.
In the GeV gamma-ray band, Fermi detected this object with a low level flux \citep{abdo_10a, abdo_10c},
though EGRET and AGILE could not \citep{hartman_99, casandjian_08, pittori_09}.

PKS$~$2155$-$304 is a nearby HBL with a redshift of $z=0.116$ \citep{falomo_93},
discovered in X-rays by the HEAO$-$1 satellite \citep{griffiths_79}.
This object has been intensively studied in VHE gamma rays,
since the Durham group first reported VHE gamma-ray emission \citep{chadwick_99}.
The CANGAROO group tried to find evidence of VHE gamma-ray emissions but without success
(e.g., \cite{roberts_99, nishijima_01, nakase_03}).
This object was confirmed as a VHE gamma-ray source by the H.E.S.S.\ group in 2005 \citep{aharonian_05a}.
In July 2006, the H.E.S.S.\ group detected an extreme flaring of PKS$~$2155$-$304,
and reported flux variation on the time scale of minutes \citep{aharonian_07},
The flare triggered CANGAROO-III observations of PKS$~$2155$-$304
and VHE gamma rays above 660\,GeV were detected at the $4.8\sigma$ significance level \citep{sakamoto_08}.
During the same period, the MAGIC group also detected PKS$~$2155$-$304 above 400\,GeV \citep{hadasch_09}.
Since then, many multiwavelength observations of PKS$~$2155$-$304 have been carried out
(e.g., \cite{aharonian_05b, aharonian_09b}),
and part of the multiwavelength campaign in 2008 (August 25 to September 6) was reported in \citet{aharonian_09a}.
CANGAROO-III observations were also conducted during the 2008 summer,
but no significant gamma-ray signal above 720\,GeV was found \citep{nishijima_09}.
The flux upper limits were obtained in this observation slightly higher than H.E.S.S. observed spectra \citep{aharonian_09a}.
This object was detected also in the GeV energy region by EGRET \citep{hartman_99, casandjian_08} and Fermi/LAT \citep{abdo_10a, abdo_10c}.

PKS$~$0537$-$441,
at a redshift of $z=0.894$ \citep{peterson_76},
is one of the most distant and the most luminous members of low-frequency peaked BL Lac (LBL) class,
and may be part of a small galaxy cluster/group with several companions \citep{heidt_03}.
Although classic/normal BL Lacs have no features in their optical spectra,
this source has several kinds of broad lines \citep{pian_02},
and while the lines disappear in high states \citep{falomo_89}.
Therefore, PKS$~$0537$-$441 is discussed as a bridging object between the
BL Lac and FSRQ classes (e.g., \cite{dammando_11, ghisellini_11}).
Although no detection of this source has been claimed in the VHE gamma-ray regime,
GeV gamma-ray detections have been reported by EGRET \citep{hartman_99, casandjian_08},
AGILE \citep{pittori_09}, and Fermi \citep{abdo_10a, abdo_10c} at high significance levels,
and GeV gamma-ray flares are often reported (e.g., \cite{tosti_08, bastieri_09, lucarelli_10, cannon_10}).

3C$~$279 is a distant FSRQ, located at a redshift of $z=0.536$ \citep{marziani_96},
and is one of the most intensively studied blazars over many years \citep{hartman_01}.
In the VHE region, the Whipple group tried to detect VHE gamma rays several times in the 1980s,
but could not \citep{cawley_85, reynolds_93}.
It was first detected in 2006 by the MAGIC group above 80\,GeV  \citep{albert_08},
and became the most distant (identified) source of VHE gamma rays.
The MAGIC group detection of the source again on 2007 January 16 \citep{berger_09},
however, observations by the H.E.S.S.\ group from January 18 to 21 did not detect this source \citep{aharonian_08}.
At GeV energies, 3C$~$279 is a well known GeV gamma-ray blazar detected by EGRET \citep{hartman_99, casandjian_08}, AGILE \citep{pittori_09}, and Fermi \citep{abdo_10a, abdo_10c},
and GeV gamma-ray flares have been reported several times (e.g., \cite{iafrate_09, hill_09}).
A multiwavelength campaign including Fermi/LAT was carried out from July 2008 to June 2009,
and a change of the optical polarization associated with a GeV gamma-ray flare was reported \citep{abdo_10b}.
During this period, the MAGIC group did not detect VHE gamma rays from this object \citep{aleksic_11}.
CANGAROO-III observations are also done during February and March 2009, and this results are reported here.

In this paper, we report the results of VHE gamma-ray searches from these four blazars,
which were observed with the CANGAROO-III telescope in the years from 2005 to 2009.
This includes a reanalysis of H$~$2356$-$309 and PKS$~$2155$-$304 data which were reported by \citet{nishijima_09}, and new results for PKS$~$0537$-$441 and 3C$~$279.
We also discuss the multiwavelength spectra using a simple one-zone leptonic model consisting of a synchrotron self-Compton (SSC) radiation and an external Compton (EC) radiation field.

\section{Instrument and observations}
The CANGAROO-III telescope system consists of four imaging atmospheric Cherenkov telescopes which are operated near Woomera,
South Australia ($136^\circ47'$E, $31^\circ06'$S, 160\,m a.s.l.).
These telescopes use altazimuth mounts and are placed at the east (T1), west (T2), south (T3), and north (T4)
corners of a diamond with sides of $\sim$100 m.
The T1 telescope, which is the oldest telescope, has not been operated for observation since 2004 due to its narrower field of view (FOV) and higher energy threshold.
We also have not operated the T2 telescope since 2008 due to severe deterioration of its mirror.
Only T3 and T4 telescopes are used for analyses in this work.
They have parabolic reflectors of 10\,m diameter with an 8\,m focal length,
and each reflector is composed of 114 small spherical mirrors of 80\,cm diameter and with an average radius of curvature of 16.4\,m,
which are made of Fiber Reinforced Plastic.
Each imaging camera is an array of 427 PMT pixels of $0.17^\circ$ size, with a total FOV of $\sim$4$^\circ$.
The PMT signals were recorded by charge ADCs and multi-hit TDCs.
Details of the mirror, the data acquisition (DAQ) system, the camera, the standard calibration method,
and the total performance of the light collecting system are given in
\citet{kawachi_01}, \citet{kubo_01}, \citet{kabuki_03}, and \citet{enomoto_06a}.

The CANGAROO-III observations of four blazars are summarized in Table \ref{obs_table}.
Each observation was made for 12--18 moonless nights using the $wobble$ mode,
in which the pointing position of each telescope was shifted in declination by $\pm0.5^\circ$ every 20 min from the target.
The wobble mode enables simultaneous observation of background regions with the target region.
These observations, except for the case of 3C$~$279, were basically carried out at zenith angles smaller than $30^\circ$.
Details of each observation follow.

The observations of H$~$2356$-$309 were carried out over 14 nights during the period between 2005 July 8 and August 12 with the three telescopes (T2, T3 and T4).
Due to DAQ trouble with the T2 telescope,
twofold coincidence data (i.e., T3 and T4) were used in the analysis.
The total observation time is 30.3 h and the mean zenith angle of the observations was 12.7$^\circ$.

The observations of PKS$~$2155$-$304 were made over 18 nights from 2008 July 29 to September 27 with two telescopes (T3 and T4)
for a total of 57.5 h,
with a mean zenith angle for the observations of 14.6$^\circ$.

PKS$~$0537$-$441 was observed in 14 nights during the period between 2008 November 24 and December 29 with the T3 and T4 telescopes.
The observations were carried out over 36.7 h and the mean zenith angle of the observations was 17.2$^\circ$.

The observations of 3C$~$279 were carried out over 12 nights during the period between 2009 February 24 and March 28 using the twofold coincidence of T3--T4.
The total observation time is 40.1 h and the mean zenith angle of the observation was 34.3$^\circ$.
Because the culmination altitude of the target position is low, the observations were carried out at zenith angles up to $\sim 60^\circ$.

\begin{table*}
\begin{center}
\caption{Summary of CANGAROO-III observations for the four blazars\label{obs_table}}
\begin{tabular}{ccccccccc}
\hline
 Blazar & \multicolumn{3}{c}{Observation periods\footnotemark[1]} & Nights\footnotemark[2] & $t_{\rm obs}$\footnotemark[3] & $\langle\theta_{\rm z}\rangle$\footnotemark[4] & \multicolumn{2}{c}{Target positions\footnotemark[5]} \\
        & Year & Begin & End & & [hrs] & [$^\circ$ ] & RA[$^\circ$ ] & Dec[$^\circ$ ] \\
\hline
H$~$2356$-$309   & 2005 & July $~$8 & Aug. 12 & 14 & 30.3 & 12.7 & 359.783 & $-$30.628  \\
PKS$~$2155$-$304 & 2008 & July 29   & Sep. 27 & 18 & 57.5 & 14.6 & 329.717 & $-$30.22   \\
PKS$~$0537$-$441 & 2008 & Nov. 24   & Dec. 29 & 14 & 36.7 & 17.2 & 84.7098 & $-$44.0858 \\
3C$~$279         & 2009 & Feb. 24   & Mar. 28 & 12 & 40.1 & 34.3\,(26.8)\footnotemark[6] & 194.046 & $-$5.789   \\
\hline
\end{tabular}
\end{center}

\footnotetext{1}{(a): Details of observation periods.}
\footnotetext{2}{(b): The number of observation nights.}
\footnotetext{3}{(c): Each observation time, in units of hour.}
\footnotetext{4}{(d): The average observation zenith angle, in units of degrees.}
\footnotetext{5}{(e): Details of target positions in J2000 coordinate.}
\footnotetext{6}{(f): Averaged zenith angle of analyzed data after zenith cut of $\theta_{\rm z} < 30^\circ$.}

\end{table*}

\section{Data reduction and analysis}

\subsection{CANGAROO-III}
Here, we briefly describe the analysis procedure for CANGAROO-III data,
which is almost identical with that given by \citet{kiuchi_09}.
More details can be found in \citet{enomoto_06a, enomoto_06b}.
After the CANGAROO-III standard calibration of the relative gain and the timing for each pixel \citep{kabuki_03},
we have cleaned the images to eliminate night sky background photons and select shower images as follows.
Each shower image is required to be a pixel cluster consisting at least five adjacent pixels,
each of which has a signal larger than 5.0~p.e.\ with an arrival time within $\pm$30 ns of the average hit timing.
Some of telescope performances depends on observational zenith angles, $\theta_{\rm z}$.
In particular energy threshold of detected gamma rays becomes higher with the zenith angle rapidly when the zenith angle exceeds $\sim$30$^\circ$.
So we cut the data of 3C$~$279 with observation during $30^\circ < \theta_{\rm z}$.
For other three objects, we use all data because more than $90\%$ of the data were taken at the zenith angle less than $30^\circ$. 
In order to reject the low quality data due to the effect of clouds and dew,
any data taken during periods when the shower rate was lower than the typical rate for clear nights were not used.
The peaks of shower rate distribution of each object are considered as the typical shower rate, 
which are 7.0, 6.8, 4.9 and 6.7\,Hz, for H$~$2356$-$309, PKS$~$2155$-$304, PKS$~$0537$-$441 and 3C$~$279, respectively.
We set shower rate thresholds to 6.0\,Hz for H$~$2356$-$309, PKS$~$2155$-$304 and 3C$~$279, and 4.5\,Hz for PKS$~$0537$-$441.
Furthermore, in order to ensure reliable arrival direction and energy estimations, i.e., to avoid the deformation of the shower images by the camera edge,
it was required that none of the brightest 15 pixels of each image should be in the outermost layer of each camera. 

The moments of the shower images ($length$ and $width$) are parameterized as defined by \citet{hillas_85}.
The arrival directions are reconstructed using those the intersection of image axes.
After the event reconstruction, the FD value is calculated from $width$ and $length$ using Fisher Discriminant (FD) method \citep{fisher_36},
which have a mathematically maximum separation between the distributions of gamma-ray events and background hadron events.

The number of detected gamma-ray events are estimated using the FD fit method \citep{enomoto_06b} for each squared angular distance bin.
As in our previous analysis \citep{sakamoto_08},
to avoid the effect of position dependence of $width$ and $length$ in the FOV, 
each of background regions was employed in an area of the opposite direction relative to the center of the FOV. 
In this analysis,
the FD distribution of background events is calculated for each ring-shaped region with area of $0.1\pi$ deg$^2$,
which radius is corresponded to the offset distance of gamma-ray analysis bin from the center of the target position.
This method cancel out the position dependence of image parameters between the region of interest and the background region via the wobble motion,
which switches each other region, and suppress the systematic difference.

The energy threshold of the integral flux, $E_{\rm th}$, is defined as the peak energy of Monte Carlo (MC) simulated gamma-ray events which  survive  pseudo hardware cuts and analysis cuts.
For differential spectra, we use three energy bands corresponding to photoelectron bands of 0--80 p.e., 80--150 p.e., and above 150 p.e., respectively.
The mean energy of each photoelectron band is used as a representative energy of the band.
The assumed VHE photon spectrum of each blazar for MC simulation is a simple power-law with the following photon indices;
$\Gamma = 3.1$ for H$~$2356$-$309, as determined by H.E.S.S.\ \citep{aharonian_06b},
$\Gamma = 3.3$ for PKS$~$2155$-$304, as reported by H.E.S.S.\ from their 2008 August to September observations \citep{aharonian_09a},
$\Gamma = 6.0$ for PKS$~$0537$-$441, as a typical value for blazar at redshift of $z \sim 0.9$ with softening by extragalactic background light (EBL) photons, because there has been no detection in VHE region,
and $\Gamma = 4.1$ for 3C$~$279, as reported by MAGIC \citep{albert_08}.

\subsection{Fermi/LAT}
In order to obtain a wide-band high-energy SED, we use non-simultaneous GeV energy data taken with the LAT on the Fermi Gamma-ray Space Telescope \citep{atwood_09}.
We analyze the data between 2008 August 5 and 2011 May 10,
using the standard analysis package of {\tt ScienceTools-v9r18p6}.

We selected the events with photon energies in the range of 0.2--300 GeV.
Two higher quality classes, the $Diffuse$ class and the $DataClean$ class,
including both $front$ and $back$ events, were used,
and the set of the instrument response functions of "P6\_V3\_DIFFUSE" was applied.
We excluded events with zenith angles larger than $105^\circ$ and time intervals when the rocking angle was larger than $52^\circ$,
so as to reduce the contamination of Earth albedo gamma rays and the bright limb of the Earth.

The standard unbinned likelihood analyses using $gtlike$ were performed for our four regions of interest (ROIs),
which are circular regions of $10^\circ$ radius centered on the positions of each target blazars, H$~$2356$-$309, PKS$~$2155$-$304, PKS$~$0537$-$441, and 3C$~$279.
We considered galactic and extragalactic diffuse emissions,
and also considered the radiation from point sources which are cataloged as 1FGL \citep{abdo_10a} within the circle regions of $15^\circ$ radius from the ROI centers.
The time variations of the sources within $10^\circ$ from the center of the ROIs were taken into account,
but for the sources outside that region the fluxes and the spectral indices are fixed at the values of the 1FGL catalog.
In addition, we found several point-like sources near the targets with a significance level of greater than at least $5\sigma$,
which are not listed in 1FGL catalog.
We summarize these unknown source positions in Table\,\ref{unknown_source},
and these are considered as point sources in our analysis.
Recently, second source catalog of Fermi/LAT (2FGL) is available \citep{abdo_11}.
Half of our additional sources might be appeared in the 2FGL catalog,
but the others have no possible counterparts.
This discrepancy might be caused by the different combinations of analyzed energy range and time range.

\begin{table*}
  \begin{center}
  \caption{Summary of additional source position of Fermi/LAT analysis (RA [$^\circ$ ], Dec [$^\circ$ ]) \label{unknown_source}}
  \begin{tabular}{ccccc}
    \hline
    ID & H$~$2356$-$309                     & PKS$~$2155$-$304                   & PKS$~$0537$-$441                  & 3C$~$279 \\
    \hline
    1  & (358.36, $-30.58$)\footnotemark[1] & (327.97, $-30.42$)\footnotemark[1] & (83.00, $-48.46$)\footnotemark[1] & (192.74, $-2.05$) \\
    2  & (359.55, $-28.89$)                 & (327.74, $-27.69$)\footnotemark[1] & (83.01, $-38.81$)\footnotemark[1] & (198.14, $-4.48$)\footnotemark[1] \\
    3  & (  3.02, $-30.58$)                 & (325.52, $-25.88$)                 & (79.67, $-45.74$)\footnotemark[1] & (189.33, $-4.39$) \\
    4  & ---                                & (335.70, $-35.00$)\footnotemark[1] & (89.03, $-43.92$)\footnotemark[1] & (187.93, $-8.47$) \\
    5  & ---                                & ---                                & ---                               & (190.67, 0.79)    \\
    \hline
  \end{tabular}
  \end{center}

  \footnotetext{1}{(a): These sources have a possible counterpart listed in 2FGL catalog.}

\end{table*}

\section{Observational results}
\begin{figure*}
  \centering
  \includegraphics[width=\linewidth]{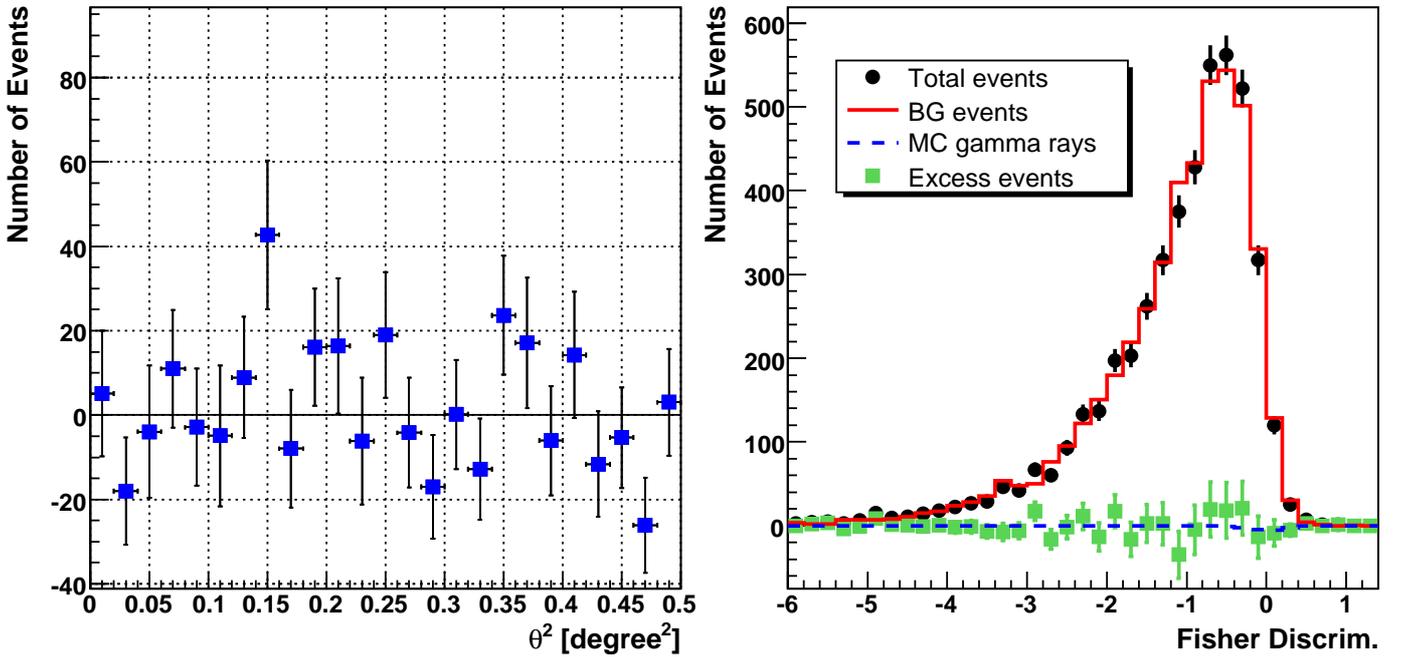}
  \caption{$\theta^2$ distribution of excess events and Fisher Discriminant distribution for H$~$2356$-$309.
     ($left$): The plots are the number of excess events as a function of the squared angular distance $\theta^2$.
     ($right$): Obtained Fisher Discriminant distributions within $\theta^2 < 0.06$.
     The black filled circles are the observed events in the ON source region,
     the red solid and the blue dashed lines are the background and the gamma-ray components estimated by the FD fit procedure.
     The green filled squares indicate gamma-ray candidates, which are the excess events between observed events and estimated background events.
  \label{theta2_h2356}}
\end{figure*}

\begin{figure*}
  \centering
  \includegraphics[width=\linewidth]{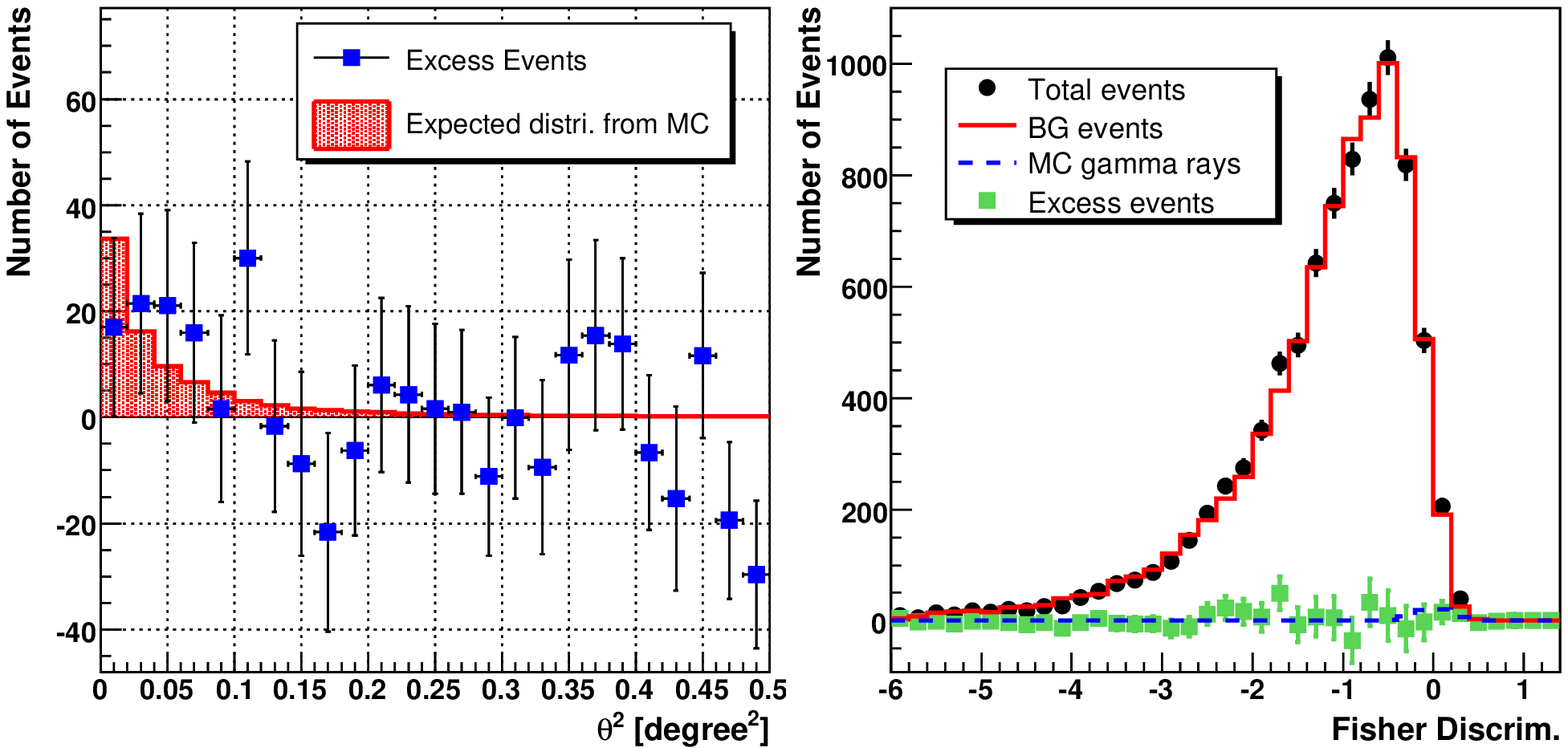}
  \caption{$\theta^2$ distribution of excess events and Fisher Discriminant distribution for PKS$~$2155$-$304.
           The panels, including symbols and lines are same as Fig. \ref{theta2_h2356}.
           The hatched histogram in the left panel represents our PSF normalized to the number of excess events.
  \label{theta2_pks2155}}
\end{figure*}

\begin{figure*}
  \centering
  \includegraphics[width=\linewidth]{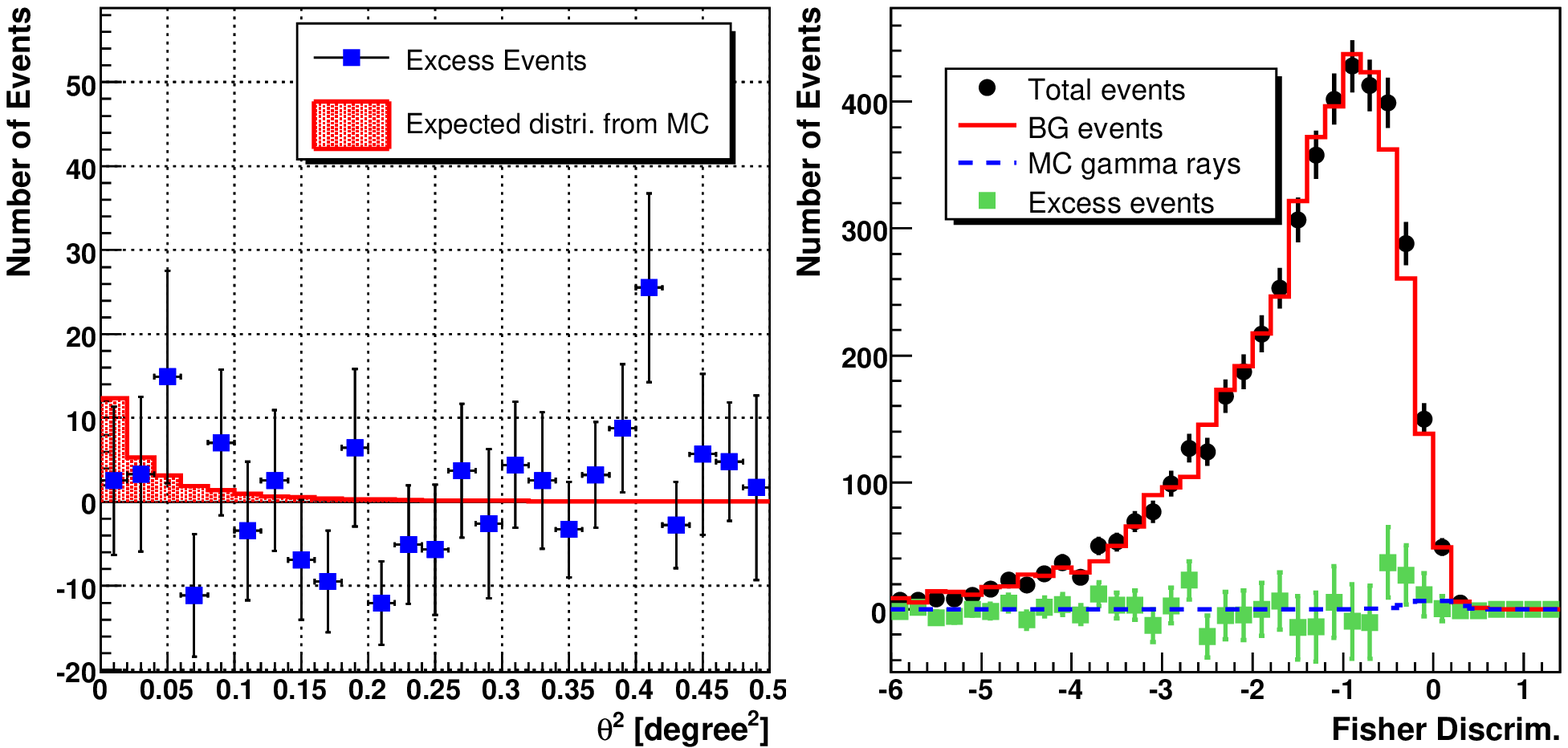}
  \caption{$\theta^2$ distribution of excess events and Fisher Discriminant distribution for PKS$~$0537$-$441.
           The panels, including symbols and lines are same as Fig. \ref{theta2_h2356}.
           The hatched histogram in the left panel represents our PSF normalized to the number of excess events.
  \label{theta2_pks0537}}
\end{figure*}

\begin{figure*}
  \centering
  \includegraphics[width=\linewidth]{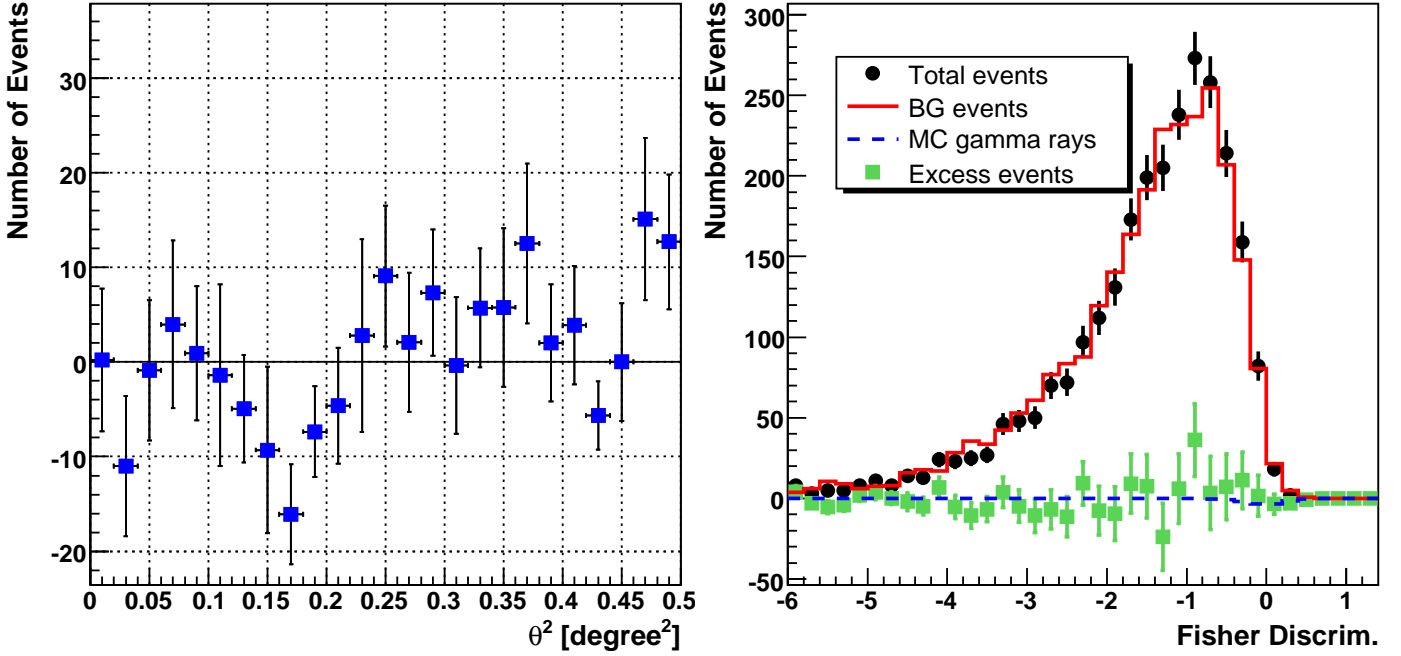}
  \caption{$\theta^2$ distribution of excess events and Fisher Discriminant distribution for 3C$~$279.
           The panels, including symbols and lines are same as Fig. \ref{theta2_h2356}.
  \label{theta2_3c279}}
\end{figure*}

After the data reduction described in previous section,
the $\theta^2$ distributions of TeV gamma-ray candidate events are obtained by fitting the FD distribution of on-source data with those for MC gamma-rays and real background data.
The left panels of Figs.\ \ref{theta2_h2356}--\ref{theta2_3c279} show results for H$~$2356$-$309, PKS$~$2155$-$304, PKS$~$0537$-$441, and 3C$~$279, respectively.
The number of excess events in each $\theta^2$ bin are plotted with $1\sigma$ statistical error bars.
The hatched histogram represents our point spread function normalized to the number of excess events in $\theta^2 < 0.06$.

The right panels of Figs. \ref{theta2_h2356}--\ref{theta2_3c279} show the Fisher Discriminant (FD) distributions.
Black filled circles with error bars are those for on-source events ($\theta^2 < 0.06$).
Red solid and blue dashed histograms are the best-fit FD distributions for background events and for MC gamma-ray events, respectively.
Green filled squares with error bars show the background-subtracted gamma-ray candidate events.

In these figures, there is no significant gamma-ray excess from the direction of any of these blazars.
We therefore calculated $2\sigma$ flux upper limits of gamma-ray emission for each object.
The results are summarized in Table \ref{table_events}.
In this table, the number of excess events, statistical significance, live time, assumed photon indices in MC simulation,
estimated energy threshold, and the integral flux upper limits are listed.

\begin{table*}
\begin{center}
  \caption{Summary of CANGAROO-III observation results \label{table_events} }
  \begin{tabular}{ccccccc}
  \hline
  Blazar & Excess\footnotemark[1] & Significance\footnotemark[2] & $t_{\rm live}$\footnotemark[3] & $\Gamma$\footnotemark[4] & $E_{\rm th}$\footnotemark[5] & $I_{\rm U.L.}(>E_{\rm th})$\footnotemark[6] \\
         & [events] & [$\sigma$] & [hrs]          &          & [TeV]        & [${\rm cm^{-2}~s^{-1}}$] \\
  \hline
  H$~$2356$-$309   & $ -16.9\pm25.2$ & $ -0.67$ & 23.3 & 3.1 & 0.68 & $<2.6\times10^{-12}$\\
  PKS$~$2155$-$304 & $~~59.5\pm29.9$ & $~~1.99$ & 41.8 & 3.3 & 0.60 & $<3.8\times10^{-12}$\\
  PKS$~$0537$-$441 & $~~20.8\pm18.0$ & $~~1.16$ & 26.6 & 6.0 & 0.51 & $<7.0\times10^{-12}$\\
  3C$~$279         & $ -11.7\pm12.9$ & $ -0.90$ & 12.5 & 4.1 & 0.72 & $<1.7\times10^{-12}$\\
  \hline
  \end{tabular}
\end{center}

\footnotetext{1}{(a): The number of gamma-ray candidates and the errors.}
\footnotetext{2}{(b): Statistical significance levels of detected gamma-ray candidates.}
\footnotetext{3}{(c): Effective exposure times.}
\footnotetext{4}{(d): Photon indices of assumed power-law spectra in MC simulations.}
\footnotetext{5}{(e): Estimated energy thresholds of integral flux.}
\footnotetext{6}{(f): $2\sigma$ upper limits of integral flux above each energy threshold.}
\end{table*}

Meanwhile in the GeV gamma-ray bands,
all of four targets are detected with high levels of significance.
Our results of H$~$2356$-$309 show a very low flux level and a flat (${\it \Gamma} = 2.0$) spectrum in the energy range of 0.9--70\,GeV.
The gamma-ray emission of PKS$~$2155$-$304 extends from 0.2 to 300\,GeV with a hard (${\it \Gamma} = 1.9$) slope.
For PKS$~$0537$-$441 and 3C$~$279, curved gamma-ray spectra are detected, with gamma-ray emission vanishing above energies of 140\,GeV and 30\,GeV, respectively.
Our gamma-ray spectra are in good agreement with the 1FGL catalog \citep{abdo_10a}.
Time variations of PKS$~$2155$-$304, PKS$~$0537$-$441 and 3C$~$279 are seen in our analysis.
Fermi/LAT GeV flux levels during CANGAROO-III observation periods show about $40\%$ higher for PKS$~$2155$-$304 and 3C$~$279, 
and about $40\%$ lower for PKS$~$0537$-$441 than long time averaged one.

\section{Spectral energy distributions and discussion}
Here, we discuss multiwavelength spectral energy distributions (SEDs) including CANGAROO-III upper limits,
GeV gamma-ray spectra of Fermi/LAT, and archival data, using a simple emission model.
Our emission model is based on the one-zone homogeneous, time independent, leptonic jet model by \citet{kataoka_99}.
In this model, we take into account the synchrotron radiation of relativistic electrons for the low frequency component,
and photons boosted by the same population of electrons via the inverse Compton scattering effect for the high frequency component.
As the seed photons, both synchrotron photons and external thermal photons are considered.
The former case is known as synchrotron self-Compton (SSC) radiation (e.g., \cite{jones_74, konigl_81}),
and the latter as external Compton (EC) radiation (e.g., \cite{dermer_93, sikora_94}).

We assume there is a spherical blob, or jet component, of radius $R$ with homogeneously tangled magnetic field $B$
which propagates at relativistic speed with a bulk Lorentz factor ${\it \Gamma}$.
The viewing angle $\theta$, that is the angle between the jet orientation and our line of sight,
is assumed to be the superluminal angle, in which case Doppler beaming factor $\delta$ can be taken $\delta={\it \Gamma}$.
In the emitting blob, injection, radiative cooling, and escaping of electrons are assumed to be in equilibrium.
We considered two models of a steady state electron energy spectrum.
One is a simple power-law spectrum expressed as
\begin{eqnarray}
  N_{\rm e}(\gamma) = Q_{0} \gamma^{p_1},
\end{eqnarray}
and the other is a smoothed broken power-law expressed as follows,
\begin{eqnarray}
  N_{\rm e}(\gamma) = \frac{Q_{0} \gamma_{\rm brk}^{p_1}}{ \left( \frac{\gamma}{\gamma_{\rm brk}} \right)^{-p_1} + \left( \frac{\gamma}{\gamma_{\rm brk}} \right)^{-p_2}},
\end{eqnarray}
where $Q_{0}$ is the normalization factor, $\gamma$ is the Lorentz factor of relativistic electrons,
$p_1$ and $p_2$ are electron spectral indices in the range from $\gamma_{\rm min}=1.0$ to $\gamma_{\rm brk}$, and $\gamma_{\rm brk}$ to $\gamma_{\rm max}$, respectively.
Due to the relativistic speed of the blob,
the density of external photons injected from behind of the blob are greatly diminished and the photons are redshifted in the blob frame.
On the contrary, external photons injected from front of the blob are strongly boosted the density and the photons are blueshifted.
Several kinds of seed photons of the EC radiation could be expected, 
such as a nuclear continuum scattered by broad-line clouds, 
line emission from broad line clouds photoionized by the central source,
and infrared photons from warm dust heated by the nucleus.
As argued by \citet{inoue_96},
for simplification of the EC radiation,
we only suppose a single component of seed photons with a blackbody spectrum of temperature $T_{\rm ext}$,
which is injected head on into the jet component.
Therefore in our SSC model,
the free parameters are $B$, $R$, $\Gamma$\,($\simeq\delta$), $Q_0$, $p_1$, $p_2$, $\gamma_{\rm brk}$, and $\gamma_{\rm max}$.
If and when we consider the EC component as well as the SSC (here after SSC + EC model),
two more free parameters, $T_{\rm ext}$, and $L_{\rm iso}$, are added,
where $L_{\rm iso}$ is an injected photon luminosity in the rest frame of the object.

Accurate estimation of the error of each physical parameter is very difficult in the case of large dimensional fitting.
As one of indications, 
we select $B$, $R$ and $\delta$, which are thought to be closely related each others,
and we estimate the $1\,\sigma$ statistical fitting errors in the three dimensional parameter space.
The other parameters of the three dimensional space are fixed to the best derived ones.

As is well known, VHE gamma rays are absorbed by Extragalactic Background Light (EBL) photons due to e$^+$e$^-$ pair creation.
There are various models of the EBL density (e.g., \cite{stecker_92, primack_99, franceschini_08}).
In this paper, we use the EBL density model by \citet{dominguez_11},
to estimate the amount of absorption of high energy gamma rays.

\subsection{H$~$2356$-$309}
\begin{figure}[tb]
  \centering
  \includegraphics[width=\linewidth]{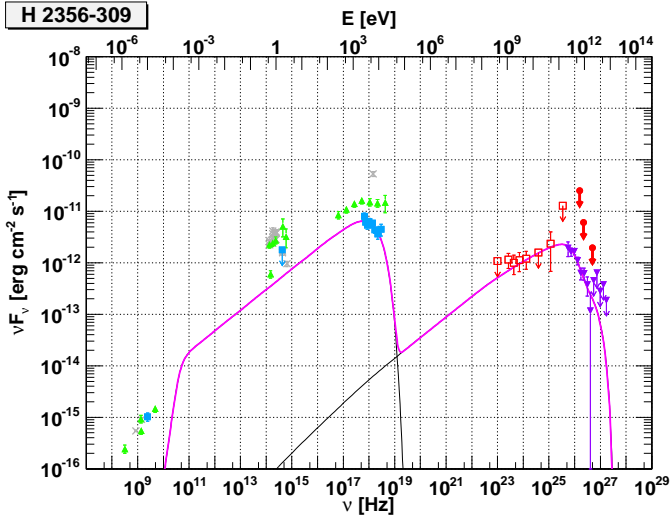}
  \caption{SEDs of H$~$2356$-$309.
     The red circles are $2\sigma$ upper limits from 2005 July to August with CANGAROO-III.
     The open red squares are an average Fermi/LAT spectrum.
     The upside-down purple triangles are the average H.E.S.S.\ spectrum of observations from 2004 to 2007 \citep{abramowski_10}.
     The filled blue squares are the 2004 RXTE campaign spectrum,
     and the green triangles are archival data during the high state in June 1998,
     both are described in \citet{aharonian_06b} and references therein.
     The light-gray crosses are archival data from the NASA/IPAC Extragalactic Database (NED).
     The bold magenta curve is the best fit to the observed SED by our SSC model.
     Two thin solid curves are the components of the synchrotron radiation and the SSC radiation.
     \label{sed_h2356}}
\end{figure}

Our SED for H$~$2356$-$309, including archival data, is shown in the Fig. \ref{sed_h2356}.
CANGAROO-III upper limits from observations in 2005 are indicated by the closed red circles,
and the average Fermi/LAT spectrum obtained by the analysis of the data
between 2008 August and 2011 May is plotted by the open red squares.

The average H.E.S.S.\ spectrum from 2004 to 2007 \citep{abramowski_10},
which includes our observation period, is indicated by the closed purple triangles.
Our upper limits have not imposed a strict limit,
however the GeV spectrum seems to connect to H.E.S.S.\ TeV flux smoothly.
No time variation of either TeV gamma-ray flux \citep{abramowski_10} or GeV gamma-ray flux from this object
was found during these observational periods.

We, then fit our SSC model to the observed spectra of gamma rays and X-rays,
which include RXTE data observed simultaneously to the H.E.S.S.\ 2004 observations \citep{aharonian_06b}.
The fit results are shown by solid lines in Fig. \ref{sed_h2356},
where the lower component shows the synchrotron photons and the higher energy component comes from SSC photons.
The observed data from X-rays to gamma rays are well reproduced.
The best fitted parameters are listed in Table \ref{table_sed_fit}.

The derived Doppler beaming factor, $\delta=59$, is larger than the previously reported values (e.g., \cite{aharonian_06b, abramowski_10, tang_10}),
and the magnetic field strength of $B=0.012$\,G is much weaker than the values so far reported (e.g., \cite{ghisellini_02, tavecchio_10}).
Fitting to the newly obtained very low flux and the flat spectrum of the GeV gamma-ray band might causes these differences.
However, derived parameters must have large uncertainty due to loosely constrained synchrotron component.
Furthermore, SSC model with a broken power-law spectrum of electrons can be also reproduce the SED,
but there is no significant improvement of the goodness of fit.
So we employ the simplest model which has no break in the electron distribution.
It may be able to determine whether the electron spectrum has a break or not if we have simultaneous soft X-ray data.
This model uncertainty is also one of large factors of parameter errors which we cannot take into account in here.

Meanwhile, neither our SSC model nor SSC + EC model could reproduce the archival flux data of the radio and the optical bands.
The radio photons are considered to be emitted in a different region from the high energy emission,
for instance, a inhomogeneous jet model \citep{tang_10} is one such model.
The optical emissions might be a combination of several components, such as a
large amount of thermal emission from the host galaxy \citep{falomo_91},
the same component as the radio emission, and the synchrotron radiation from relativistic electrons.

\subsection{PKS$~$2155$-$304}
\begin{figure}[tb]
  \centering
  \includegraphics[width=\linewidth]{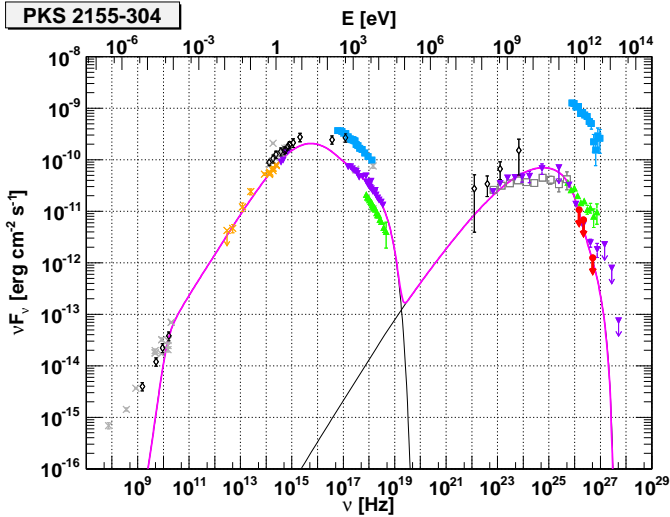}
  \caption{SEDs of PKS$~$2155$-$304.
     The red circles are $2\sigma$ upper limits from 2008 July to September with CANGAROO-III.
     The open squares are an average Fermi/LAT spectrum.
     The purple inverted triangles are simultaneous spectra of H.E.S.S., Fermi, RXTE, and ATOM in 2008 \citep{aharonian_09a}.
     The blue filled squares are the exceptional high state in 2006 July  \citep{aharonian_09b}.
     The green triangles indicate the 2003 RXTE--H.E.S.S.\ campaign results \citep{aharonian_05b}.
     The open diamonds are archival campaign data from 1994 November \citep{vestrand_95}.
     The light-gray crosses are archival data of NED.
     The orange crosses are also NED archival data but the data was used to SED fittings.
     The bold magenta curve is the best fit to the observed SED by our SSC model.
     Two thin solid curves indicate the synchrotron and the SSC radiation.
     \label{sed_pks2155}}
\end{figure}

The wide-band SED of PKS$~$2155$-$304, observed at various times, is summarized in Fig. \ref{sed_pks2155}.
The closed red circles indicate the upper limits calculated from the data taken in 2008 with CANGAROO-III.
The Fermi/LAT data are also plotted by the gray open squares.
Our flux upper limits in TeV energies are lower than the flux we observed in 2006 \citep{sakamoto_08},
and also lower than the reported fluxes by the H.E.S.S.\ team during multiwavelength campaign period in 2003 and 2008 
which are plotted by the green triangles and purple inverted triangles, respectively.
Although our observation period includes their 2008 campaign period,
the results are not necessarily inconsistent with each other as PKS$~$2155$-$304
is one of the most variable blazars in the VHE gamma-ray region.

Our radiation model is fitted to the data from the infrared to TeV gamma-ray energies,
which includes archival infrared and optical data from NED and the simultaneous broad band spectrum observed in the
2008 campaign period reported by \citet{aharonian_09a}.
The report include Fermi/LAT results,
and the GeV gamma-ray activity in this period is slightly higher than the average one which we obtained from the data between 2008 and 2011,
though the TeV gamma-ray fluxes show a quiescent state.
Our SSC model with single power-law electron spectrum cannot reproduce the SED,
and hard photon slope of soft X-ray band require the break of spectrum strongly.
The solid line in Fig. \ref{sed_pks2155} shows our SSC model prediction of the multiwavelength SED assuming a broken power-law spectrum of electrons.
The obtained physical parameters are summarized in Table \ref{table_sed_fit}.
PKS$~$2155$-$304 has been well studied and a lot of effort was invested to understand its SED particularly by one-zone SSC model (e.g., \cite{aharonian_05b, foschini_08, ghisellini_08, katarzynski_08, kusunose_08}).
Our model fits the synchrotron component very well,
and the inverse Compton component is also well fitted.
A large beaming factor $\delta=43$ is obtained,
which is a little bit larger than the previously reported values.
Other parameters are in good agreement with already reported ones,
although some of the models assume a double-broken power-law spectrum as an electron energy distribution.

\subsection{PKS$~$0537$-$441}
\begin{figure}[tb]
  \centering
  \includegraphics[width=\linewidth]{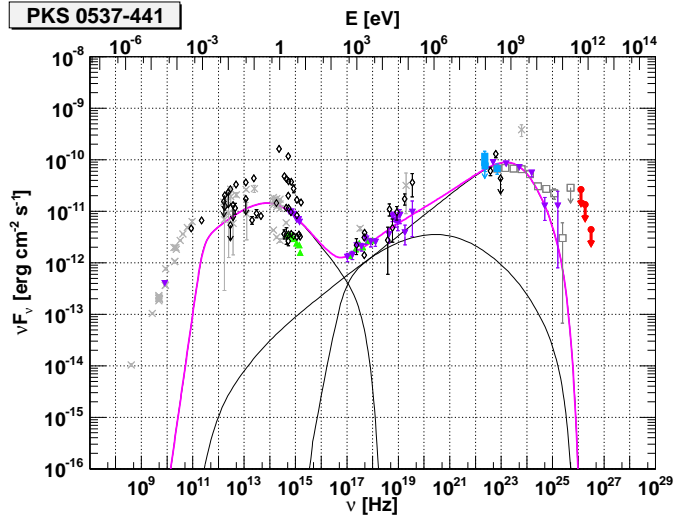}
  \caption{SEDs of PKS$~$0537$-$441.
     The bold red circles with arrows are $2\sigma$ upper limits in 2008 November and December with CANGAROO-III.
     The open squares are an average Fermi/LAT spectrum.
     The purple upside-down triangles correspond to the quasi-simultaneous radio to gamma-ray spectra in August to October of 2008 \citep{abdo_10d}.
     The blue closed squares are AGILE results in 2008 October  \citep{pucella_10}, and first AGILE catalog \citep{pittori_09}.
     The green triangles are Swift results during 2008 August to October \citep{ghisellini_10}.
     The open diamonds are 2005 Swift and REM three simultaneous observations and archival data \citep{pian_07}.
     The light-gray crosses are archival data from NED.
     The best fitted curve to the SED by our SSC + EC model is indicated by the bold magenta curve.
     Individually, three thin solid curves are the synchrotron, the SSC, and the EC components, from the lower to the higher energy peaks, respectively.
     \label{sed_pks0537}}
\end{figure}

The observed SEDs of PKS$~$0537$-$441 at several epochs are compiled in Fig. \ref{sed_pks0537}.
Our flux upper limits obtained from the observations in 2008 are the only data at TeV energies,
which are plotted by the closed red circles in this figure,
together with our analyzed GeV gamma-ray data of Fermi/LAT symbolized by the open gray squares.
The quasi-simultaneous multiwavelength data observed in 2008 assembled by \citet{abdo_10d},
the data from AGILE and Swift, and archival data from NED are also plotted in the same figure.

We apply our radiation model to fit the data taken in 2008,
where our averaged GeV data were not used and the data from the campaign periods was used instead,
because we have no data of radio to X-ray bands simultaneous to our CANGAROO-III observations.
The simple one-zone SSC model fails to fit the observed SED well,
so we take into account another component, the EC component, and the results are shown in Fig. \ref{sed_pks0537},
where the low-frequency component, mid-frequency component, and high-frequency component are the
synchrotron radiation, SSC radiation, and EC radiation, respectively.
From this figure, it is found that contributions from SSC photons and EC photons to the X-rays are roughly comparable,
while the gamma-ray photons mainly originate from the EC mechanism.
The physical values obtained by the fitting are listed in Table \ref{table_sed_fit}.
Although the strength of the magnetic field obtained here is much weaker than the typical values reported before by others (e.g., \cite{pian_02, ghisellini_10, pian_07}),
some models of this object proposed by \citet{pucella_10}, 
which use a similar state of multiwavelength behavior with ours,
predict a magnetic field as weak as the one we derived.
From our model, the size of emitting blob is derived to be $R=4.1\times10^{16}$ cm,
and the bulk Lorentz factor is obtained to be ${\it \Gamma}=24$,
which are consistent with other works \citep{pian_02, ghisellini_10, pian_07, pucella_10}.
Although the best of the black body temperature of EC seed photon is derived to be $T_{\rm ext}=2.3\times10^3$\,K,
it is difficult to infer the physical situations.
Because this parameter is one of the most insensitive parameter in our model.

It is seen from the figure that our SSC + EC model cannot reproduce the observed high radio fluxes
which requires a huge size of the emission blob in our model and is
implausible from the time scales of the observed flux variations \citep{heidt_96}.
The bulk of the radio emission therefore probably originates much farther out along the jet.

\subsection{3C$~$279}
\begin{figure}[tb]
  \centering
  \includegraphics[width=\linewidth]{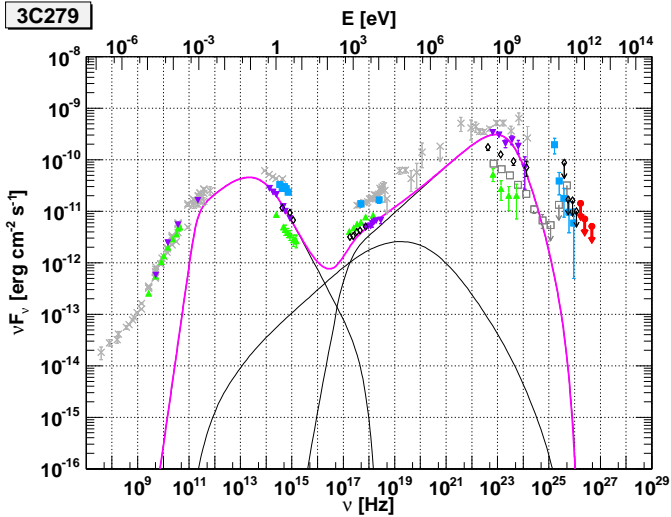}
  \caption{SEDs of 3C$~$279.
     The red circles are $2\sigma$ upper limits in 2009 February to March from CANGAROO-III.
     The open squares are an average spectrum of Fermi/LAT.
     The purple inverted triangles and green triangles are simultaneous multiwavelength spectra in the epochs of MJD 54880 to 54885 and MJD 54950 to 54960, respectively \citep{abdo_10b}.
     The blue closed squares indicate a large gamma-ray flare state of 2006 February 23,
     and the open diamonds indicate the spectrum in the epoch of 2009 January 21 to February 1 \citep{albert_08, aleksic_11}.
     The light-gray crosses are archival data from NED.
     The bold magenta curve is the best fit to the SED by our SSC + EC model.
     There are three thin solid curves, from the lower to the higher frequency components, the synchrotron, the SSC, and the EC radiation.
     \label{sed_3c279}}
\end{figure}

Many simultaneous multiwavelength campaigns of 3C$~$279 have been performed to date.
The SEDs obtained from some of those observations for wide ranges of frequencies are shown in Fig. \ref{sed_3c279}.
CANGAROO-III upper limits are plotted by the red closed circles.
Open gray squares indicate GeV gamma-ray spectrum from our analysis of Fermi/LAT data.
One of the simultaneous observation campaigns was conducted in February 2009,
which data are symbolized by the purple inverted triangles in this figure, just before our observations with CANGAROO-III.
In this period, 3C$~$279 maintains an active state in GeV band with flux levels  clearly higher than the average flux,
while in the X-ray band it is at a relatively low flux level.

For 3C$~$279, the SSC + EC model fits the observed SEDs from optical to gamma-ray bands well,
where simultaneous data taken in campaign periods of 2009 are used for fitting.
The model spectra are represented by the solid line in Fig. \ref{sed_3c279},
and the best set of model parameters is summarized in Table \ref{table_sed_fit}.
In this model, boosted EC photons are the main origin of gamma rays,
and only a small fraction of SSC radiation contributes to the X-ray bands.

Although in our model, the luminosity of the SSC component is relatively low compared to earlier works (e.g., \cite{aleksic_11, ghisellini_10, collmar_10, inoue_96, moderski_03}),
most of our parameters such as the size of moving blob $R = 1.0\times10^{17}$\,cm,
the bulk Lorentz factor $\Gamma = 19$, and the strength of the magnetic field $B = 0.56$\,G, are roughly consistent with them.
The lowness of SSC luminosity is caused by a complex combination of the blob size $R$, the magnetic field $B$, and the electron density $n_{\rm e}$.
The obtained seed photon luminosity that is required from EC component photons is $L_{\rm iso} = 2.2\times10^{41}$ erg s$^{-1}$.
If it is assumed for the accretion disk luminosity that the UV luminosity of $L = 2\times10^{45}$ erg s$^{-1}$ estimated from the observations in the historical lowest state of 3C$~$279 \citep{pian_99},
it could be probable that $L_{\rm iso} = 2.2\times10^{41}$ erg s$^{-1}$ corresponds to 0.01 \% of the disk luminosity.
The temperature of seed photons spectrum of EC radiation is derived to be $T_{\rm ext}=5.7\times10^3$\,K,
but it is difficult to discuss its origin because this parameter is one of the most insensitive parameter to the SED.

As in the case of the other three objects,
the flux of the synchrotron component in this model is insufficient to explain the large observed radio flux.
The possibility that the radio emission regions in blazars might be much farther from the central engine has been discussed (e.g., \cite{hughes_91, carrara_93}).

\begin{table*}
\begin{center}
\caption{Summary of the best parameter sets of the SED modelings }
\label{table_sed_fit}
\begin{tabular}{lccccc}
  \hline
                        & [unit]       & H$~$2356$-$309        & PKS$~$2155$-$304     & PKS$~$0537$-$441     & 3C$~$279 \\
  class                 &              & HBL                   & HBL                  & LBL                  & FSRQ     \\
  \hline
  $z$ &                 & 0.165                & 0.116                & 0.894                & 0.536    \\
  $R$ & [cm]            & $\left(5.8_{-1.6}^{+1.4}\right) \times 10^{15}$ & $\left(5.8_{-0.2}^{+0.9}\right) \times 10^{16}$ & $\left(4.1_{-0.5}^{+0.6}\right) \times 10^{16}$ & $\left(1.0_{-0.1}^{+0.1}\right) \times 10^{17}$ \\
  $B$ & [G]             & $\left(1.2_{-0.7}^{+1.1}\right) \times 10^{-2}$ & $\left(2.7_{-0.1}^{+0.7}\right) \times 10^{-2}$ & $\left(4.7_{-0.6}^{+0.6}\right) \times 10^{-1}$ & $\left(5.6_{-0.5}^{+0.6}\right) \times 10^{-1}$ \\
  $\Gamma$              &              & $59_{-23}^{+60}$      & $43_{-8.6}^{+2.3}$   & $24_{-0.7}^{+0.7}$   & $19_{-0.8}^{+0.8}$   \\
  $\theta_{\rm v}$      & [$^\circ$ ]  & 0.97                  & 1.3                  & 2.4                  & 3.0                  \\
  \hline
  $Q_{0}$               & [cm$^{-3}~\gamma^{-1}$] & $1.5 \times 10^{4}$   & $8.2 \times 10^{-1}$ & $1.5 \times 10^{4}$  & $1.4 \times 10^{3}$  \\
  $p_1$                 &              & $-2.2$                & $-1.4$               & $-2.2$               & $-2.1$               \\
  $p_2$                 &              & ---                   & $-3.9$               & $-4.3$               & $-4.9$               \\
  $\gamma_{\rm min}$    &              & 1.0                   & 1.0                  & 1.0                  & 1.0                  \\
  $\gamma_{\rm brk}$    &              & ---                   & $2.6 \times 10^{4}$  & $2.3 \times 10^{3}$  & $1.2 \times 10^{3}$  \\
  $\gamma_{\rm max}$    &              & $8.1 \times 10^{5}$   & $8.4 \times 10^{5}$  & $8.7 \times 10^{4}$  & $8.7 \times 10^{4}$  \\
  \hline
  $L_{\rm iso}$         & [erg s$^{-1}$]    & ---                   & ---                  & $1.5 \times 10^{40}$ & $2.2 \times 10^{41}$ \\
  $T_{\rm ext}$         & [K]               & ---                  & ---                  & $2.3 \times 10^{3}$  & $5.7 \times 10^{3}$  \\
  \hline
  $N_{\rm total}$       &              & $1.0 \times 10^{52}$ & $1.6 \times 10^{51}$ & $3.6 \times 10^{54}$ & $5.3 \times 10^{54}$ \\
  $n_{\rm e}$           & [cm$^{-3}$]  & $1.3 \times 10^{4}$  & $2.0 \times 10^{0}$  & $1.2 \times 10^{4}$  & $1.3 \times 10^{3}$  \\
  $\varepsilon_{\rm e}$ & [erg cm$^{-3}$]   & $5.7 \times 10^{-2}$ & $5.5 \times 10^{-4}$ & $4.8 \times 10^{-2}$ & $5.8 \times 10^{-3}$ \\
  \hline
\end{tabular}
\end{center}

\footnotetext{}{The rows from upper to lower are as follows: blazar name, the class, $z$: the redshift, $R$: the size of radiative component with $1\,\sigma$ statistical fitting error, 
  $B$: the magnetic field strength with $1\,\sigma$ statistical fitting error,
  $\Gamma$: the bulk Lorentz factor of the radiative component with $1\,\sigma$ statistical fitting error, 
  $\theta_{\rm v}$: the angle between the jet orientation and our line of sight, $Q_{0}$: the normalization factor of electron spectrum, $p_1$, $p_2$: the electron spectral indices at low and high energies, 
  $\gamma_{\rm min}$, $\gamma_{\rm brk}$, $\gamma_{\rm max}$: the minimum, the break, and the maximum of electron Lorentz factors, $L_{\rm iso}$: the seed photon luminosity of EC radiation in blazar frame, 
  $T_{\rm ext}$: the blackbody temperature of seed photon spectrum of EC radiation, $N_{\rm total}$: the total number of electrons, $n_{\rm e}$: the electron number density, and $\varepsilon_{\rm e}$: the electron energy density, 
  respectively.}
\end{table*}

\section{Conclusions}
We have observed four selected blazars, H$~$2356$-$309, PKS$~$2155$-$304, PKS$~$0537$-$441 and 3C$~$279, with the CANGAROO-III imaging atmospheric Cherenkov telescope from 2005 to 2009.
No statistically significant excess of events above 510--720\,GeV from the direction of any of these objects was found,
and we derived flux upper limits for VHE gamma-ray emissions.
In addition, we analyzed GeV gamma-ray data between 0.2\,GeV and 300\,GeV taken with Fermi/LAT from August 2008 to May 2011.

To derive some important physical parameters of these blazars,
we consider a simple leptonic jet model to explain the multiwavelength SEDs including GeV and TeV spectra, even though non-simultaneous.
The observed SED of H$~$2356$-$309 (HBL) could be explained by a simple SSC model with a single power-law electron spectrum,
and to keep a consistency with GeV spectrum,
we need to assume a large beaming factor $\delta = 59$ and weak magnetic field strength of 0.012\,G.
Radiation from PKS$~$2155$-$304, a nearby HBL,
was well modeled by the SSC scenario, and obtained parameters are consistent with earlier works.
PKS$~$0537$-$441, a luminous LBL,
was studied and we found SSC + EC model could explain the observed multiwavelength spectrum
where the EC component is dominant in the gamma-ray photons.
The SED of one of the distant FSRQ, 3C$~$279, were also well explained by the SSC + EC model.

Additionally, from our parameter fit results in Table \ref{table_sed_fit} as the HBL to the FSRQ of blazar sub-classes, 
it is seen that the beaming factor becomes smaller,
and in contrast the strength of the magnetic field becomes stronger and the size of blob becomes larger.
Although we could not take into account of the uncertainties such as simultaneity of the data and the differences of models,
these latter two tendencies are in agreement with the proposed blazar sequence (e.g., \cite{fossati_98, ghisellini_98}).


\section*{Acknowledgments}
This work is supported by a Grant-in-Aid for Scientific Research by the Japan Ministry of Education,
Culture, Sports, Science and Technology, the Australian Research Council, JSPS Research Fellowships,
and Inter-University Researches Program by the Institute for Cosmic Ray Research.
We many thank the Defense Support Center Woomera and BAE Systems.

\end{document}